\documentstyle[11pt,aaspp4]{article}
\input psfig

\begin{document}

\title{Jet directions in Seyfert galaxies: B and I imaging data}

\author
  {H. R. Schmitt\altaffilmark{1,2,3,4,6}, A. L. Kinney\altaffilmark{1,2,3,5}}
\altaffiltext{1}{Space Telescope Science Institute, 3700 San Martin Drive, 
Baltimore, MD 21218, USA}
\altaffiltext{2}{Visiting Astronomer Cerro Tololo Interamerican Observatory,
National Optical Astronomy Observatories, which is operated by AURA, Inc.
under a cooperative agreement with the National Science Foundation}
\altaffiltext{3}{Visiting Astronomer Kitt Peak National Observatory
National Optical Astronomy Observatories, which is operated by AURA, Inc.
under a cooperative agreement with the National Science Foundation}
\altaffiltext{4}{Visiting Astronomer Lick Observatory, operated by the
University of California Observatories}
\altaffiltext{5}{Present address: NASA Headquarters, 300 E St., Washington, DC20546}
\altaffiltext{6}{email:schmitt@stsci.edu}

\date{\today}

\begin{abstract}

We present the results of broad-band B and I imaging observations for a
sample of 88 Seyfert galaxies (29 Seyfert 1's and 59 Seyfert 2's),
selected from a mostly isotropic property, the flux at 60$\mu$m. We
also present the B and I imaging results for an additional sample of 20
Seyfert galaxies (7 Seyfert 1's and 13 Seyfert 2's),
selected from the literature and known to have extended radio
emission.  The I band images are fitted with ellipses to determine the
position angle and ellipticity of the host galaxy major axis. This
information will be used in a future paper, combined with information
from radio observations, to study the orientation of radio jets
relative to the plane of their host galaxies (Kinney et al. 2000). Here we
present surface brightness profiles and magnitudes in the B and I bands,
as well as mean ellipticities and major axis position angles.

\end{abstract}


\section {Introduction}

We have recently shown (Schmitt et al. 1997; Clarke, Kinney \& Pringle
1998; see also Nagar \& Wilson 1999) that there is no correlation
between the position angle of radio jets and disk major axes
in Seyfert galaxies, confirming previous results based on smaller
samples (Ulvestad \& Wilson 1984; Brindle et al. 1990; Baum et al.
1993).  Clarke et al. (1998) and Nagar \& Wilson (1999) showed, using a
statistical inversion technique, that the observed values of $\delta$ (the
difference between the position angle of the jet and the host galaxy
disk major axis) and $i$ (inclination of the galaxy disk relative to
the line of sight) can be reproduced by a homogeneous distribution
of angles $\beta$ between the jet and the galaxy disk axis.

These results contradict the expectation that the jets should be
aligned perpendicular to the galaxy disk. The simplest assumption
about the feeding of the accretion disk and the black hole
suggests that the gas comes from the host galaxy disk, so it is
natural to expect both disks to be aligned and have the same angular
momentum vector. Since jets are emitted perpendicular to the accretion
disk, we would expect them to be aligned with the host galaxy
minor axis, which is not observed. These studies give us information
about the inner workings of Seyferts and may shed some light on the
processes involved in the feeding of the AGN.

Although the results from Clarke et al. (1998) and Nagar \& Wilson
(1999) were statistically significant, they had two major limitations,
their samples and most of their measurements were obtained from the
literature. This indicates that their results could be biased by
selection effects, like the preferential selection of galaxies which
have jets shining into the plane of the galaxy, resulting in brighter
radio emission and narrow line regions, which would be easier to
detect. From the point of view of the data, using measurements
collected from the literature can also influence the results, since
different authors are likely to measure the position angle of radio jets,
the disk inclination and the position angle of the host galaxy
major axis using different techniques and data of different quality.

In order to improve the data relative to previous studies, we obtained radio
continuum maps at 3.6cm, optical broad band images and spectroscopy for
a sample of Seyfert galaxies selected from a mostly isotropic property,
the flux at 60$\mu$m. In this way we avoid selection effects and create
a homogeneous database, with measurements done using a consistent
technique.

Another possible improvement which will be used in the analysis paper
(Kinney et al. 2000), is the distinction between which side of the
galaxy minor axis is closer to Earth. According to Clarke et al.
(1998), this information can improve the statistical determination of
the $\beta$-distribution by a factor of 2. One way to obtain this
information is from the inspection of dust lanes in the galaxies'
images. Dust lanes can be seen in the near side of the galaxy, because
they are highlighted against the background bulge light. Due to this
fact, we decided to obtain images in the B and I bands, with a large
wavelength separation, which will allow us to search for dust lanes.
Another way to obtain this information is from the rotation curve of
the galaxy. Knowing which side of the galaxy is approaching us and
assuming that the spiral arms are trailing, we can determine which side
of the minor axis is closer to Earth. In order to do this, we obtained
long-slit spectra, with the slit aligned close to the host galaxy major
axis, for several objects in our sample.

In this paper we present the broad-band B and I imaging data. The radio
continuum observations and optical spectroscopic data will be presented
elsewhere. In Section 2 we present the samples used in our study. The
description of the observations and reductions is given in Section 3,
and the measurements are presented in Section 4. A summary is given in
Section 5.

\section{Sample}
\label{sample}

\subsection{60$\mu$m sample}

In order to avoid selection effects as much as possible, we have chosen a
sample from a mostly isotropic property, the flux at 60$\mu$m.
According to the torus models of Pier \& Krolik (1992), which are the
most anisotropic and hence the most conservative models, the
circumnuclear torus radiates nearly isotropically at 60$\mu$m.

Our sample includes 88 Seyfert galaxies (29 Seyfert 1's and 59 Seyfert
2's), which correspond to all galaxies from the de Grijp et al. (1987,
1992) sample of warm IRAS galaxies with redshift z$\leq0.031$. The
galaxies in this sample were selected based on the quality of the
60$\mu$m flux, Galactic latitude $|b|>20^{\circ}$, and
25$\mu$m$-60\mu$m color in the range $-1.5<\alpha(25/60)<0$, chosen to
exclude starburst galaxies as much as possible. The candidate AGN
galaxies were all observed spectroscopically (de Grijp et al. 1992) to
confirm their activity class as being Seyfert 1 or Seyfert 2 and {\it
not} a lower level of activity such as starburst or LINER. The distance
limit of z$\leq0.031$ is large enough to encompass a statistically
significant number of objects yet close enough to ensure that radio
features can be resolved.

Table 1 presents the galaxies in the de Grijp et al.  (1987) catalog,
selected for our study. We list their catalog numbers, names,
coordinates, the total exposure times in the B and I bands, and the
observing runs in which the galaxies were observed.

\subsection{Additional sample}

Parts of the study presented by Kinney et al. (2000) will also use an
additional sample of 53 Seyfert galaxies selected from the literature.
This sample comprises Seyferts known to have extended radio emission,
used in previous studies (such as Schmitt et al. 1997; or Nagar et al.
1999) but which are not in the 60$\mu$m sample. For 20 of these
galaxies (7 Seyfert 1's and 13 Seyfert 2's) we were able to obtain B
and/or I images during our observing runs. Table 2 gives the names of
the galaxies, their coordinates, total exposure times in B and I bands
and the observing run in which they were observed.

Some of the galaxies in the additional sample were used in previous
papers, but we now consider that they should not be included in this
analysis. The reasons to exclude them are the fact that they are in
interacting systems, mergers, or the radio emission is not extended
enough to allow a reliable measurement of the position angle of the
jet. For these galaxies,  Column 8 (Comments) of Table 2 gives the
reasons why they are excluded.

\section{Observations and reductions}
\label{obs}

The data presented in this paper were obtained in 5 different observing
runs, using 3 different observatories. The dates of these observing
runs, corresponding telescopes and instruments are shown in Table 3.

The CTIO observations were done in the 0.9m telescope with
focal ratio f/13.5, using the detector T2K6 for run $a$ and
detector T2K3 for run $d$.  Both CCD's have the same plate scale,
which gives a pixel size of 0.384$^{\prime\prime} pixel^{-1}$. The
images in run $a$ were obtained using the whole CCD area of
2048$\times$2048 pixels, reading it out using 4 different amplifiers,
which gives a field of view of $\approx13^{\prime}\times13^{\prime}$.
For run $d$ we used only a 1024$\times$1024 section of the CCD, reading
it out using one amplifier, which gives a field of view of
$\approx7.5^{\prime}\times7.5^{\prime}$.

The observations at Lick Observatory were done in the 1.0m Nickel
telescope with focal ratio f/17, using Dewar \#5 in both runs ($b$ and
$c$), which gives a pixel size of 0.248$^{\prime\prime} pixel^{-1}$.
We used the whole CCD area (1024$\times$1024 pixels) for these
observations, which gives a field of view of
$\approx4.8^{\prime}\times4.8^{\prime}$.  The KPNO observations were
done in the 0.9m telescope with focal ratio f/7.5, using the detector
T2KA, which gives a pixel size of 0.688$^{\prime\prime} pixel^{-1}$. We
used only a 1024$\times$1024 section of the CCD, which gives a field of
view of $\approx11.7^{\prime}\times11.7^{\prime}$.

We followed the same observing procedure for each one of the runs. For
each night we obtained a series of bias images (between 20 and 50
exposures), dome flats (between 15 and 30 exposures per filter) and sky
flats (between 5 and 10 exposures per filter). We did not obtain dark
images, because our exposure times were short enough that the
contribution of dark current was negligible.

The reductions were done following standard IRAF procedures. The
individual images were overscanned, bias subtracted and divided by the
normalized flat field. Tests showed that, for each observing run, there
was no significant differences between calibration frames from
individual nights. Therefore, all frames were combined and we used the
resulting images, which had a higher S/N, for the data reduction. The
images were flat-fielded using only the sky flats, since tests showed
that the dome flats had inhomogeneous illumination.

To calibrate the images in the Cousins system, each night we observed
several standard star fields from Graham (1982) and Landolt (1992). We
estimate that the photometric accuracy of our observations is of the
order of 0.05 mag. To avoid the saturation of the nuclear region and to
eliminate cosmetic defects and cosmic rays, the images were dithered
using 3 or more exposures of 400s or less.  For three of the galaxies
in the 60$\mu$m sample it was possible to obtain images in only one of
the bands (I for MRK1040 and B for IRAS16382-0613 and UGC10683B).
Furthermore, runs $a$ and $e$ took place close to full moon,
resulting in shallower B images.

Since our images will be used to compare the position angles in the
radio and optical, it is important to determine the orientation of the
CCD's relative to the equatorial plane. This was done using stars in
the images, which showed that the fields are not rotated. The final
orientation of the images is N up and E to the left, with an
uncertainty of $\approx1^{\circ}$.

\section{Measurements}

In Figure 1 we present the I band images of the galaxies, organized
following the same order of Tables 1 and 2. In the case of UGC10683B
and IRAS16382-0613, for which we were not able to obtain I band images,
the B band images are shown instead.

Measurements of disk ellipticities and major axis position angles
(defined as the angle measured from N to E) were
obtained fitting ellipses to the isophotes of the galaxies, using the
routine ``ellipse'' in the STSDAS package of IRAF. We have chosen to
fit the ellipses over the I band images because they were deeper, and
also because this band is more sensitive to old stars, so the outer
isophotes are not disturbed by HII regions like they can be in the B
images.

The ellipses were fitted from the inner $\approx0.7^{\prime\prime}$ of
the I images, out to the level where the surface brightness reached the
3$\sigma$ level above the background (this limiting value is listed in
Tables 4 and 5).  The background level and its standard deviation
($\sigma$) were determined from several blank regions around the
galaxy. For some galaxies, with bright stars close to the low surface
brightness isophotes (e.g. NGC3783), the ellipse fitting procedure was
truncated before it reached the 3$\sigma$ level, to avoid the
disturbance of the fit by these stars. Ellipses centers were hold fixed
at the nuclear position, and were fitted using a constant increment of
the semi-major axis, 2 pixels for runs {\it a, b, c} and $d$ and 1
pixel for run $e$, which corresponds to $\approx0.77^{\prime\prime}$
for runs $a$ and $d$, $\approx0.5^{\prime\prime}$ for runs $b$ and $c$
and $\approx0.69^{\prime\prime}$ for run $e$.  The surface brightness
of the 3$\sigma$ level in I was typically 21-22.5 mag~arcsec$^{-2}$,
which corresponds to 23-24.5 mag~arcsec$^{-2}$ in the B band, assuming
that the mean color of spiral disks is (B-I)$\approx2$ (H\'eraudeau,
Simien \& Mamon 1996; see also our own measurements in Figure 2).

The ellipse parameters obtained from the fit of the I band images were
used to measure the surface brightness profile of the B images, thus
allowing a direct comparison between the two measurements.  In Figure 2
we show the surface brightness profiles, major axis position angle
(PA$_{MA}$) and disk ellipticity (e), defined as e=1-b/a, where b/a is
the ratio between the minor and major axis. Notice that the ellipse
parameters in the inner 1$^{\prime\prime}-2^{\prime\prime}$ are
unreliable, because they were made on scales smaller or comparable to
the seeing.

In Tables 4 and 5, for the 60$\mu$m and additional samples,
respectively, we present the size of the ellipse major axis at the
3$\sigma$ level above the background, the surface brightness of this
level, the integrated magnitude inside this region and the seeing
during the observations, for both B and I bands. The integrated
magnitudes were calculated by integrating the flux inside the ellipses
corresponding to the 3$\sigma$ isophote, using the major axis lengths
given in Tables 4 and 5, the ellipticities and PA$_{MA}$'s given in Tables
6 and 7 for the 60$\mu$m and additional samples, respectively. Notice
that, since there is a difference between the 3$\sigma$ level of the B
and I band images, the B images are usually shallower and not as
extended as the I images, the ellipse parameters used to measure
the integrated magnitudes of these two bands are slightly different.
This is the reason why Tables 6 and 7 give different values of
ellipticity and PA$_{MA}$ for B and I bands.

The I band ellipticities and PA$_{MA}$'s were obtained by averaging the
results from the ellipses fitted between the isophotes 3$\sigma$ and
4$\sigma$ above the background (4 to 10 points, depending on the
galaxy).  We adopted this procedure to eliminate large spurious
variations, since these values will be combined with radio measurements
to study the orientation of radio jets relative to the host galaxy disk
axis in these galaxies (Kinney et al. 2000).  An inspection of the
radial profiles in Figure 2 shows that there is not too much variation
in the ellipticities and PA$_{MA}$'s of the outer isophotes, with the
exception usually being the galaxies close to face-on and interacting
systems.

\section{Summary}

We presented B and I band images for a sample of 88 Seyfert galaxies
selected from a mostly isotropic property, the flux at 60$\mu$m, as
well as for an additional 20 Seyfert galaxies with extended radio
emission. The isophotes of the I band images were fitted with ellipses
to determine the surface brightness profiles, the
ellipticities and position angles of the host galaxy major axis. The
parameters obtained with these fits were used to measure the surface
brightness profiles in the B band. These images were also used to
measure the integrated B and I magnitudes of the galaxies. These
measurements will be combined with information from radio observations
to study the orientation of radio jets relative to the host galaxy disk
(Kinney et al. 2000).

\acknowledgements 

We would like to acknowledge the hospitality and help from the staff at
CTIO, KPNO and Lick Observatories during the observations. We also
would like to thank Blaise Canzian for useful comments on the
measurement of ellipticities and position angles in spiral galaxies,
as well as the anonymous referee for helpful comments.
This work was supported by NASA grants NAGW-3757, AR-5810.01-94A,
AR-6389.01-94A and the HST Director Discretionary fund D0001.82223.
This research made use of the NASA/IPAC Extragalactic Database (NED),
which is operated by the Jep Propulsion Laboratory, Caltech, under
contract with NASA. We also used the Digitized Sky Survey, which was
produced at the Space telescope Science Institute under U.S. Government
grant NAGW-2166.

\begin{figure}
\caption{I band images of the 60$\mu$m sample and additional
sample.  In the case of UGC10683B and IRAS16382-0613, we present the B
band images, because we were not able to observe them in the I band.
The orientation of the images is N to the top and E to the left. The
names of the galaxies are shown on the top left corner of the image
which are organized following their order of appearence in Tables 1 and
2. The grayscale is a liner stretch from 0 counts (the background level) to
the 19 mag~arcsec$^{-2}$ surface brightness level. The contours start
at 3$\sigma$ above the background level, presented in Tables 4 and 5,
and increase in powers of 2, times 3$\sigma$ (3$\sigma\times2^n$).}
\end{figure}

\begin{figure}
\caption{B (open triangles) and I (filled squares) surface brightness
(in units of mag~arcsec$^{-2}$), position angle of the major axis
(PA$_{MA}$) and ellipticity (e) as a function of the semi-major axis.
The position angles and ellipticities were obtained fitting ellipses to
the I band images. The B surface brightness were measured using the
ellipse parameters obtained from the I band images, which allow a
direct comparison between the measurements in the two bands. The errorbars
represent $\pm 1\sigma$ (the standard deviation).}
\end{figure}


\begin{references}

Baum, S.A., O'Dea, C.P., de Bruyn, A.G., \& Pedlar, A. 1993, ApJ, 419,
553

Brindle, C., Hough. J.H., Bailey, J.A., Axon, D.J., Ward, M.J.,
Sparks, W.B., \& McLean, I.S. 1990, MNRAS, 244, 577

Clarke, C.J., Kinney, A.L., \& Pringle, J.E. 1998, ApJ, 495, 189

De Grijp, M.H.K., Keel, W. C., Miley, G.K., Goudfrooij, P. \& Lub, J. 1992,
A\&AS, 96, 389

De Grijp, M.H.K., Miley, G.K., \& Lub, J. 1987, A\&AS, 70, 95

Graham, J. A. 1982, PASP, 94, 244

H\'eraudeau, P., Simien, F. \& Mamon, G. A. 1996, A\&AS, 117, 417

Kinney, A.L., Schmitt, H.R., Clarke C.J., Pringle, J.E., Ulvestad, J.S.,
\& Antonucci, R.R.J. 2000, ApJ, submitted

Landolt, A. U. 1992, AJ, 104, 340

Nagar, N.M., \& Wilson, A.S. 1999, ApJ, 516, 97

Schmitt, H.R., Kinney, A.L., Storchi-Bergman, T., \& Antonucci,
R. 1997, ApJ, 477, 623

Ulvestad, J.S., \& Wilson, A.S. 1984, ApJ, 285, 439

\end{references}
\end{document}

\begin{small}
\begin{deluxetable}{rlcccrrl}
\tablewidth{0pc}
\tablecaption{The Data for the 60$\mu$m Sample}
\tablehead{\colhead{Number}&\colhead{Name}&\colhead{Type}&\colhead{RA(2000)}&
\colhead{Dec(2000)}&\colhead{B}& \colhead{I}&\colhead{Run}\cr
&&&&&(s)&(s)&}
\startdata
16 &MRK\,348		&2& 00 48 47.1& $+$31 57 25& 1200 & 1200 & b\cr
24 &TOL\,0109$-$38	&2& 01 11 27.5& $-$38 05 01& 720  & 720  & a\cr
26 &MRK\,1    		&2& 01 16 07.2& $+$33 05 22& 1800 & 1200 & c,b\cr
27 &MRK\,359  		&1& 01 27 32.5& $+$19 10 44& 960  & 960  & a\cr
33 &MRK\,573  		&2& 01 43 57.8& $+$02 21 00& 720  & 720  & a\cr
37 &IRAS\,01475$-$0740 	&2& 01 50 02.7& $-$07 25 48& 1600 & 2000 & a\cr
41 &ESO\,153$-$G20 	&2& 02 06 03.6& $-$55 11 35& 900  & 1200 & a\cr
47 &MRK\,1040		&1& 02 28 14.5& $+$31 18 42&  --  & 1200 & b\cr
52 &ESO\,355$-$G25 	&2& 02 31 51.0& $-$36 40 16& 1200 & 1200 & a\cr
53 &UGC\,2024 		&2& 02 33 01.2& $+$00 25 15& 900  & 900  & a\cr
57 &NGC\,1068 		&2& 02 42 40.7& $-$00 00 48& 720  & 720  & a\cr
67 &MCG$-$02$-$08$-$039 &2& 03 00 29.8& $-$11 24 59& 900  & 900  & a\cr
68 &UGC\,2514   	&1& 03 03 48.5& $-$01 06 13& 1200 & 1200 & a\cr
75 &IRAS\,03106$-$0254 	&2& 03 13 08.3& $-$02 43 19& 1200 & 1200 & a\cr
78 &IRAS\,03125$+$0119 	&2& 03 15 05.3& $+$01 30 30& 1200 & 1200 & a\cr
83 &MRK\,607 		&2& 03 24 48.7& $-$03 02 33& 1600 & 1600 & a\cr
85 &ESO\,116$-$G18	&2& 03 24 53.1& $-$60 44 20& 1600 & 1200 & a\cr
141&IRAS\, 04385$-$0828	&2& 04 40 54.9& $-$08 22 22& 1600 & 1200 & a\cr
154&IRAS\, 04502$-$0317	&2& 04 52 44.0& $-$03 13 01& 1600 & 1600 & a\cr
156&IRAS\, 04507$+$0358	&2& 04 53 24.9& $+$04 03 39& 900  & 1600 & d,a\cr
157&ESO\,33$-$G02 	&2& 04 55 59.6& $-$75 32 27& 1600 & 1600 & a\cr
174&MCG$-$05$-$13$-$017 &1& 05 19 35.5& $-$32 39 30& 1600 & 1200 & a\cr
196&MRK\,3	 	&2& 06 15 36.3& $+$71 02 15& 900  & 1200 & e\cr
203&UGC\,3478     	&1& 06 32 47.3& $+$63 40 25& 900  & 1200 & e\cr
209&MRK\,6	 	&1& 06 52 12.3& $+$74 25 37& 1200 & 1200 & e\cr
213&FAIRALL\,265	&1& 06 56 29.7& $-$65 33 43& 1600 & 1600 & a\cr
225&MRK\,79	  	&1& 07 42 32.8& $+$49 48 35& 1200 & 1600 & b\cr
227&MRK\,10	   	&1& 07 47 29.1& $+$60 56 01& 900  & 1200 & b-e,e\cr
233&UGC\,4155	   	&1& 08 00 20.4& $+$26 36 56& 1200 & 1200 & e\cr
236&MRK\,622	   	&2& 08 07 41.0& $+$39 00 15& 1200 & 1200 & e\cr
244&ESO\,18$-$G09	&2& 08 24 07.4& $-$77 46 53& 900  & 900  & d\cr
253&MCG$-$01$-$24$-$012	&2& 09 20 51.7& $-$08 04 47& 900  & 900  & d\cr
260&MRK\,1239 		&1& 09 52 19.1& $-$01 36 43& 900  & 900  & d\cr
272&NGC\,3393		&2& 10 48 24.0& $-$25 09 40& 900  & 900  & d\cr
278&NGC\,3516 		&1& 11 06 47.5& $+$72 34 07& 1200 & 1200 & e\cr
281&IRAS\,11215$-$2806 	&2& 11 24 02.6& $-$28 23 15& 900  & 900  & d\cr
282&MCG$-$05$-$27$-$013	&2& 11 27 23.3& $-$29 15 31& 900  & 1200 & d\cr
283&MRK\,176     	&2& 11 32 35.3& $+$52 56 50& 1200 & 1200 & e\cr
286&NGC\,3783      	&1& 11 39 01.8& $-$37 44 20& 900  & 900  & d\cr
292&MRK\,766    	&1& 12 18 26.5& $+$29 48 46& 1200 & 1200 & e\cr
293&NGC\,4388   	&2& 12 25 46.7& $+$12 39 41& 1200 & 1200 & e\cr
299&NGC\,4507	    	&2& 12 35 37.0& $-$39 54 31& 900  & 900  & d\cr
301&NGC\,4593	  	&1& 12 39 39.4& $-$05 20 39& 900  & 900  & d\cr
302&TOL\,1238$-$364    	&2& 12 40 52.9& $-$36 45 22& 900  & 900  & d\cr
306&NGC\,4704	    	&2& 12 48 46.4& $+$41 55 16& 1200 & 1200 & e\cr
309&MCG$-$02$-$33$-$034	&1& 12 52 12.4& $-$13 24 54& 900  & 900  & d\cr
310&ESO\,323$-$G32 	&2& 12 53 19.8& $-$41 38 14& 900  & 900  & d\cr
313&MCG$-$04$-$31$-$030	&2& 13 07 06.0& $-$23 40 43& 900  & 900  & d\cr
314&IRAS\,13059$-$2407 	&2& 13 08 42.1& $-$24 23 00& 900  & 900  & d\cr
317&MCG$-$03$-$34$-$064	&2& 13 22 24.4& $-$16 43 43& 900  & 900  & d\cr
322&ESO\,383$-$G18	&2& 13 33 26.3& $-$34 00 59& 900  & 900  & d\cr
324&MCG$-$6$-$30$-$15	&1& 13 35 53.7& $-$34 17 45& 900  & 900  & d\cr
329&NGC\,5347	 	&2& 13 53 17.8& $+$33 29 27& 900  & 1200 & e\cr
340&IRAS\,14082$+$1347	&2& 14 10 41.4& $+$13 33 29& 1200 & 1200 & e\cr
341&NGC\,5506	   	&2& 14 13 14.8& $-$03 12 27& 900  & 1200 & e\cr
344&NGC\,5548	   	&1& 14 17 59.5& $+$25 08 12& 900  & 1200 & e\cr
349&IRAS\,14317$-$3237	&2& 14 34 44.9& $-$32 50 28& 900  & 900  & d\cr
354&IRAS\,14434$+$2714	&2& 14 45 36.8& $+$27 02 05& 1200 & 1200 & e\cr
369&UGC\,9826	   	&1& 15 21 33.0& $+$39 12 01& 900  & 1200 & e\cr
377&UGC\,9944	   	&2& 15 35 47.8& $+$73 27 02& 1200 & 1200 & e\cr
383&IRAS\,15480$-$0344	&2& 15 50 41.5& $-$03 53 18& 1200 & 1200 & e\cr
409&IRAS\,16288$+$3929	&2& 16 30 32.6& $+$39 23 03& 1200 & 1200 & e\cr
418&IRAS\,16382$-$0613	&2& 16 40 52.3& $-$06 18 52& 1200 &  --  & a-e\cr
445&UGC\,10889 	  	&2& 17 30 20.7& $+$59 38 20& 1200 & 900  & e\cr
447&MCG$+$03$-$45$-$003	&2& 17 35 32.7& $+$20 47 48& 1200 & 1200 & e\cr
471&FAIRALL\,49	  	&2& 18 36 58.1& $-$59 24 09& 1200 & 720  & a\cr
473&FAIRALL\,51	  	&1& 18 44 54.3& $-$62 21 49& 1200 & 1500 & a\cr
497&ESO\,143$-$G09	&1& 20 08 46.1& $-$61 05 56& 900  & 900  & a\tablebreak
501&FAIRALL\,341	&2& 20 19 59.0& $-$52 37 20& 1600 & 1200 & a\cr
510&UGC\,11630	  	&2& 20 47 33.5& $+$00 24 42& 720  & 720  & a\cr
512&PKS\,2048$-$57	&2& 20 52 02.0& $-$57 04 09& 900  & 900  & a\cr
530&NGC\,7213	  	&1& 22 09 16.2& $-$47 10 00& 900  & 900  & a\cr
537&MRK\,915	  	&1& 22 36 46.5& $-$12 32 43& 1200 & 1200 & a\cr
538&UGC\,12138	  	&1& 22 40 17.0& $+$08 03 14& 1600 & 1600 & a\cr
540&AKN\,564	  	&1& 22 42 39.3& $+$29 43 31& 1200 & 1200 & b\cr
549&UGC\,12348	  	&2& 23 05 19.4& $+$00 11 28& 1200 & 1600 & a\cr
555&NGC\,7674	  	&2& 23 27 56.7& $+$08 46 45& 1200 & 1200 & a\cr
590&MRK\,590		&1& 02 14 33.6& $-$00 46 00& 1600 & 1600 & a\cr
594&MRK\,1058		&2& 02 49 51.8& $+$34 59 17& 1800 & 1200 & c,b\cr
602&NGC\,1386		&2& 03 36 45.4& $-$35 59 57& 1200 & 1200 & a\cr
615&MCG\,+8-11-11	&1& 05 54 53.6& $+$46 26 22& 1200 & 1200 & b\cr
627&MRK\,705		&1& 09 26 03.3& $+$12 44 04& 1200 & 1200 & e\cr
634&NGC\,3281		&2& 10 31 52.1& $-$32 51 13& 900  & 900  & d\cr
638&UGC\,6100		&2& 11 01 34.0& $+$45 39 14& 1200 & 1200 & e\cr
665&NGC\,4941		&2& 13 04 13.1& $-$05 33 06& 900  & 900  & d\cr
703&UGC\,10683B		&1& 17 05 00.4& $-$01 32 29& 1200 &  --  & e\cr
708&ESO\,103-G35	&2& 18 38 20.3& $-$65 25 42& 720  & 720  & a\cr
721&NGC\,7212		&2& 22 07 02.0& $+$10 14 00& 1200 & 1200 & a\cr
\tablenotetext{}{Column 1 shows the galaxy number in the de Grijp et
al. (1987) catalogue, Column 2 shows their names and Column 3 the
Seyfert type.  Columns 4 and 5 give the Right Ascention and
Declination, respectively. Columns 6 and 7 give the total exposure time
of the B and I images, respectively.  Column 8 shows the observing run
when the galaxies were imaged, according to the code given in Table 1.}
\enddata
\end{deluxetable}
\end{small}

\clearpage
\newpage

\begin{small}
\begin{deluxetable}{lcccrrll}
\tablewidth{0pc}
\tablecaption{The Additional Sample}
\tablehead{\colhead{Name}&\colhead{Type}&\colhead{RA(2000)}&
\colhead{Dec(2000)}&\colhead{B}& \colhead{I}&\colhead{Run}&\colhead{Comments}\cr
&&&&(s)&(s)&&}
\startdata
NGC\,235A		&1& 00 42 52.8& $-$23 32 29&  --  &  --  & --&interacting\cr
NGC\,513		&2& 01 24 26.9& $+$33 47 58&  --  &  --  & --&\cr
NGC\,526A		&1& 01 23 54.2& $-$35 03 55&  --  &  --  & --&interacting\cr
NGC\,591		&2& 01 33 31.3& $+$35 40 05&  --  &  --  & --&\cr
NGC\,788		&2& 02 01 06.4& $-$06 48 56&  --  &  --  & --&not extended in radio\cr
NGC\,1144		&2& 02 55 12.2& $-$00 11 01&  --  &  --  & --&merger\cr
ESO\,417-G6		&2& 02 56 21.5& $-$32 11 06&  --  &  --  & --&not extended in radio\cr
MRK\,1066		&2& 02 59 58.6& $+$36 49 14&  --  &  --  & --&\cr
NGC\,1365		&1& 03 33 36.4& $-$36 08 25&  --  &  --  & --&\cr
MRK\,618		&1& 04 36 22.2& $-$10 22 34&  --  &  --  & --&\cr
ESO\,362-G8		&2& 05 11 09.0& $-$34 23 36& 900  & 900  & d &\cr
NGC\,2110		&2& 05 52 11.4& $-$07 27 22&  --  &  --  & --&\cr
NGC\,2273		&2& 06 50 08.7& $+$60 50 45& 800  & 1200 & e &\cr
ESO\,428-G14		&2& 07 14 32.9& $-$29 14 04& 900  & 900  & d &\cr
MRK\,78			&2& 07 42 41.7& $+$65 10 37& 1200 & 1200 & e &\cr
NGC\,2622		&1& 08 38 10.9& $+$24 53 43& 1200 & 1200 & e &\cr
NGC\,2639		&1& 08 43 38.0& $+$50 12 20& 1200 & 1200 & e &\cr
MRK\,110		&1& 09 25 12.9& $+$52 17 11&  --  &  --  & --&merger\cr
UGC\,5101		&1& 09 35 51.6& $+$61 21 11& 1200 & 800  & e &merger\cr
NGC\,2992		&2& 09 45 42.0& $-$14 19 35&  --  & 900  & d &interacting\cr
NGC\,3081		&2& 09 59 29.5& $-$22 49 35&  --  &  --  & --&not extended in radio\cr
NGC\,3227		&1& 10 23 30.6& $+$19 51 54&  --  & 1600 & e &\cr
MRK\,34			&2& 10 34 08.6& $+$60 01 52&  --  & 900  & e &\cr
NGC\,3362		&2& 10 44 51.7& $+$06 35 48&  --  & 900  & e &interacting\cr
MRK\,423		&2& 11 26 48.5& $+$35 15 03&  --  & 900  & e &merger\cr
NGC\,4051		&1& 12 03 09.6& $+$44 31 53&  --  &  --  & --&\cr
NGC\,4074		&2& 12 04 29.7& $+$20 18 59&  --  &  --  & --&not extended in radio\cr
NGC\,4117		&2& 12 07 46.1& $+$43 07 35&  --  &  --  & --&\cr
NGC\,4151		&1& 12 10 32.6& $+$39 24 21&  --  &  --  & --&\cr
MRK\,231		&1& 12 56 14.2& $+$56 52 25&  --  &  --  & --&merger\cr
ESO\,323-G77		&1& 13 06 26.6& $-$40 24 50&  --  &  --  & --&\cr
NGC\,5135		&2& 13 25 44.0& $-$29 50 02&  --  &  --  & --&\cr
NGC\,5252		&2& 13 38 16.0& $+$04 32 33&  --  & 1200 & e &\cr
MRK\,266		&2& 13 38 17.7& $+$48 16 34&  --  &  --  & --&merger\cr
MRK\,268		&2& 13 41 11.1& $+$30 22 41&  --  & 1200 & e &\cr
MRK\,270		&2& 13 41 05.8& $+$67 40 20& 1200 & 1200 & e &\cr
NGC\,5273		&1& 13 42 08.3& $+$35 39 15& 1200 & 1200 & e &\cr
MRK\,273		&2& 13 44 42.1& $+$55 53 13&  --  &  --  & --&merger\cr
IC\,4329A		&1& 13 49 19.3& $-$30 18 34&  --  & 720  & d &\cr
MRK\,279		&1& 13 53 03.4& $+$69 18 30&  --  &  --  & --&\cr
MRK\,463E		&2& 13 56 02.9& $+$18 22 19&  --  &  --  & --&merger\cr
NGC\,5643		&2& 14 32 40.7& $-$44 10 29&  --  &  --  & --&\cr
NGC\,5728		&2& 14 42 23.9& $-$17 15 11&  --  &  --  & --&\cr
NGC\,5929		&2& 15 26 06.2& $+$41 40 14& 1200 & 1200 & e &interacting\cr
ESO\,137-G34		&2& 16 35 14.2& $-$58 04 41&  --  &  --  & --&\cr
MRK\,509		&1& 20 44 09.7& $-$10 43 25&  --  &  --  & --&\cr
NGC\,7172		&2& 22 02 01.7& $-$31 52 18&  --  &  --  & --&\cr
IC\,5169		&2& 22 10 10.0& $-$36 05 18&  --  &  --  & --&\cr
NGC\,7450		&1& 23 00 47.8& $-$12 55 07&  --  & 1200 & b &\cr
NGC\,7465		&2& 23 02 00.9& $+$15 57 53&  --  &  --  & --&\cr
MRK\,926		&1& 23 04 43.5& $-$08 41 08&  --  &  --  & --&\cr
NGC\,7672		&2& 23 27 31.4& $+$12 23 06&  --  & 900  & b &\cr
NGC\,7743		&2& 23 44 21.5& $+$09 56 05&  --  &  --  & --&\cr
\tablenotetext{}{Column 1 gives the galaxy name and Column 2 the Seyfert type.
Columns 3 and 4 give the Right Ascention and Declination, respectively.
Columns 5 and 6 give the total exposure time of the B and I images,
respectively.  Column 7 shows the observing run when the galaxies were
imaged, according to the code given in Table 1. For those galaxies which will
not be used in future analysis, we give in Column 8 the reason to exclude them.}
\enddata
\end{deluxetable}
\end{small}

\clearpage
\newpage

\begin{small}
\begin{deluxetable}{llrlrrl}
\tablewidth{0pc}
\tablecaption{Observing Runs}
\tablehead{\colhead{Date}&\colhead{Telescope}&\colhead{f-ratio}&
\colhead{Instrument}&\colhead{FOV}&\colhead{Scale}&\colhead{Code}\cr
&&&&$^{\prime}$&$^{\prime\prime}/pixel$}
\startdata
October  03-09/10 1998 &CTIO 0.9m&13.5 &CCD Direct T2K6    &13.0&0.384  &a\cr
November 13-18/19 1998 &Lick 1.0m&17.0 &CCD Direct CCD \#5 & 4.8&0.248  &b\cr
December 17-21/22 1998 &Lick 1.0m&17.0 &CCD Direct CCD \#5 & 4.8&0.248  &c\cr
February 10-12/13 1999 &CTIO 0.9m&13.5 &CCD Direct T2K3    & 7.5&0.384  &d\cr
April    23-29/30 1999 &KPNO 0.9m& 7.5 &CCD Direct T2KA    &11.7&0.688  &e\cr
\enddata
\end{deluxetable}
\end{small}

\clearpage
\newpage

\begin{small}
\begin{deluxetable}{lrrrrrrrr}
\tablewidth{0pc}
\tablecaption{Magnitudes of the 60$\mu$m Sample}
\tablehead{\colhead{Name}&\colhead{Aperture B}&
\colhead{3$\sigma$B}&\colhead{B}&\colhead{(seeing B)}&\colhead{Aperture I}& 
\colhead{3$\sigma$I}&\colhead{I}&\colhead{(seeing I)}\cr
&(arcsec)&(mag arcsec$^{-2}$)&(mag)&(arcsec)&
(arcsec)&(mag arcsec$^{-2}$)&(mag)&(arcsec)}
\startdata
MRK\,348		&39.7  &23.80 &14.96 &0.9 &42.7  &22.56  &12.87 & 0.9\cr
TOL\,0109$-$38		&46.1  &22.18 &13.84 &1.2 &84.5  &22.13  &11.67 & 1.0\cr
MRK\,1    		&27.8  &24.04 &15.62 &2.6 &31.7  &22.69  &13.52 & 1.1\cr
MRK\,359  		&23.0  &22.67 &14.44 &1.4 &32.3  &21.65  &12.57 & 1.2\cr
MRK\,573  		&24.6  &23.08 &14.66 &1.2 &36.9  &21.92  &12.43 & 1.0\cr
IRAS\,01475$-$0740 	&16.9  &24.52 &16.58 &2.7 &20.0  &23.03  &14.46 & 1.2\cr
ESO\,153$-$G20 		&30.7  &22.40 &15.09 &1.3 &76.8  &22.35  &12.10 & 1.1\cr
MRK\,1040		&---   &---   &---   &--- &123.0 &22.56  &11.73 & 1.5\cr
ESO\,355$-$G25 		&49.2  &23.60 &14.00 &2.6 &49.2  &21.65  &12.02 & 1.0\cr
UGC\,2024 		&26.1  &23.19 &15.26 &1.3 &43.0  &22.26  &13.02 & 1.2\cr
NGC\,1068 		&192.0 &22.84 &9.93  &1.5 &318.7 &21.86  &7.73  & 1.3\cr
MCG$-$02$-$08$-$039 	&33.8  &23.07 &15.57 &1.2 &56.8  &21.99  &12.80 & 1.3\cr
UGC\,2514   		&32.3  &22.79 &14.96 &1.5 &92.2  &22.14  &11.74 & 1.2\cr
IRAS\,03106$-$0254 	&20.0  &22.61 &16.37 &1.5 &58.4  &22.45  &12.86 & 1.2\cr
IRAS\,03125$+$0119 	&15.4  &22.90 &15.87 &1.4 &32.3  &22.16  &13.46 & 1.4\cr
MRK\,607 		&81.4  &23.60 &13.79 &2.7 &101.4 &22.45  &11.41 & 1.5\cr
ESO\,116$-$G18		&36.9  &23.36 &15.36 &2.5 &33.8  &20.54  &12.85 & 1.3\cr
IRAS\, 04385$-$0828	&18.4  &22.57 &16.19 &1.0 &35.3  &22.15  &13.52 & 1.0\cr
IRAS\, 04502$-$0317	&20.0  &23.16 &15.92 &2.5 &30.7  &22.64  &13.65 & 1.2\cr
IRAS\, 04507$+$0358	&36.9  &24.18 &15.25 &1.6 &27.7  &21.79  &13.02 & 1.2\cr
ESO\,33$-$G02 		&30.7  &23.58 &15.13 &2.5 &43.0  &21.83  &12.58 & 1.2\cr
MCG$-$05$-$13$-$017 	&43.0  &23.59 &13.60 &1.7 &44.5  &22.08  &11.94 & 1.2\cr
MRK\,3	 		&53.7  &23.99 &13.94 &1.7 &82.6  &22.29  &11.28 & 1.3\cr
UGC\,3478     		&62.8  &23.43 &14.11 &1.6 &111.5 &22.98  &11.83 & 1.9\cr
MRK\,6	 		&111.5 &22.98 &11.83 &1.8 &67.4  &22.77  &12.22 & 1.3\cr
FAIRALL\,265		&29.2  &23.48 &14.99 &2.0 &46.1  &22.29  &13.11 & 1.1\cr
MRK\,79	  		&47.6  &23.91 &14.39 &1.1 &77.4  &22.76  &12.10 & 1.0\cr
MRK\,10	   		&64.7  &23.67 &14.45 &1.8 &92.2  &23.15  &12.43 & 1.3\cr
UGC\,4155	   	&30.3  &23.41 &14.92 &2.4 &45.4  &21.87  &12.60 & 1.9\cr
MRK\,622	   	&27.5  &23.10 &14.96 &1.8 &30.3  &22.45  &13.40 & 2.1\cr
ESO\,18$-$G09		&49.2  &24.17 &14.45 &1.3 &45.3  &22.07  &12.88 & 1.3\cr
MCG$-$01$-$24$-$012	&66.1  &24.38 &15.00 &1.3 &61.4  &22.60  &13.16 & 1.1\cr
MRK\,1239 		&21.5  &24.41 &15.19 &1.3 &23.8  &22.75  &13.21 & 1.4\cr
NGC\,3393		&92.2  &24.16 &13.19 &1.3 &90.6  &22.11  &11.04 & 0.9\cr
NGC\,3516 		&83.9  &23.83 &12.78 &1.9 &103.2 &22.13  &10.68 & 1.9\cr
IRAS\,11215$-$2806 	&30.7  &23.85 &15.58 &1.1 &30.7  &21.53  &13.43 & 1.1\cr
MCG$-$05$-$27$-$013	&79.9  &24.50 &14.96 &1.1 &92.2  &23.01  &12.27 & 1.0\cr
MRK\,176     		&31.7  &23.50 &15.18 &1.8 &41.3  &22.28  &13.21 & 1.6\cr
NGC\,3783     	 	&106.0 &24.31 &12.62 &1.3 &86.8  &22.08  &10.77 & 1.0\cr
MRK\,766    		&46.8  &23.60 &13.96 &1.7 &55.0  &22.15  &12.08 & 1.7\cr
NGC\,4388   		&227.0 &23.66 &12.11 &1.8 &192.6 &21.90  &10.38 & 1.7\cr
NGC\,4507	    	&76.8  &24.02 &13.11 &1.1 &69.1  &21.83  &11.12 & 0.9\cr
NGC\,4593	  	&179.7 &24.25 &12.12 &1.3 &216.6 &22.88  &9.97  & 1.3\cr
TOL\,1238$-$364    	&61.4  &23.94 &13.03 &1.3 &61.4  &22.46  &11.52 & 1.7\cr
NGC\,4704	    	&45.4  &23.91 &14.73 &1.9 &56.4  &22.76  &12.68 & 1.6\cr
MCG$-$02$-$33$-$034	&96.0  &23.90 &13.71 &1.1 &96.0  &22.39  &11.73 & 1.3\cr
ESO\,323$-$G32 		&58.4  &24.15 &13.85 &1.3 &73.7  &22.56  &11.53 & 1.1\cr
MCG$-$04$-$31$-$030	&87.6  &24.23 &14.24 &1.3 &115.2 &22.81  &11.70 & 1.0\cr
IRAS\,13059$-$2407 	&32.3  &24.27 &16.55 &1.1 &41.5  &23.20  &14.01 & 0.9\cr
MCG$-$03$-$34$-$064	&46.9  &24.13 &14.44 &0.9 &73.7  &22.73  &11.83 & 1.1\cr
ESO\,383$-$G18		&27.7  &23.26 &15.63 &1.4 &30.7  &22.08  &13.92 & 1.5\cr
MCG$-$6$-$30$-$15	&44.5  &23.95 &14.11 &1.1 &50.7  &22.41  &11.87 & 1.1\cr
NGC\,5347	 	&79.8  &23.90 &13.61 &1.7 &100.4 &22.99  &11.58 & 1.4\cr
IRAS\,14082$+$1347	&26.1  &23.19 &15.64 &1.4 &38.5  &22.25  &13.42 & 1.3\cr
NGC\,5506	   	&148.6 &23.62 &13.14 &1.7 &166.5 &22.33  &10.94 & 1.5\cr
NGC\,5548	   	&99.1  &24.79 &13.04 &1.7 &112.8 &23.47  &11.43 & 1.5\cr
IRAS\,14317$-$3237	&23.0  &24.05 &16.33 &1.0 &30.7  &22.69  &13.91 & 1.2\cr
IRAS\,14434$+$2714	&27.5  &24.30 &15.73 &1.6 &26.1  &22.38  &14.02 & 1.5\cr
UGC\,9826	   	&55.0  &24.39 &15.11 &1.6 &66.1  &23.00  &13.02 & 1.1\cr
UGC\,9944	   	&63.3  &23.65 &14.99 &1.5 &78.4  &22.74  &12.77 & 1.2\cr
IRAS\,15480$-$0344	&20.6  &24.32 &16.32 &1.3 &37.2  &22.98  &13.45 & 1.4\cr
IRAS\,16288$+$3929	&30.3  &23.93 &15.62 &1.2 &39.9  &22.80  &13.55 & 1.6\cr
IRAS\,16382$-$0613	&19.3  &23.32 &16.60 &1.3 &---   &---    &---   &--- \cr
UGC\,10889 	  	&74.3  &25.54 &14.97 &1.2 &71.6  &22.95  &12.88 & 2.3\cr
MCG$+$03$-$45$-$003	&35.8  &24.20 &15.71 &1.2 &41.3  &21.44  &13.30 & 2.0\cr
FAIRALL\,49	  	&15.4  &22.77 &15.61 &1.5 &27.7  &22.01  &12.85 & 1.3\cr
FAIRALL\,51	  	&33.8  &22.88 &14.74 &1.7 &79.9  &22.29  &12.14 & 1.2\cr
ESO\,143$-$G09		&39.9  &22.42 &13.89 &1.8 &61.4  &22.35  &11.66 & 1.3\tablebreak
FAIRALL\,341		&58.4  &24.60 &14.41 &2.3 &61.4  &22.65  &12.17 & 1.0\cr
UGC\,11630	  	&29.2  &22.10 &14.60 &1.1 &78.3  &21.91  &11.79 & 1.0\cr
PKS\,2048$-$57		&47.7  &22.81 &13.75 &1.5 &93.7  &21.68  &10.80 & 1.5\cr
NGC\,7213	  	&69.1  &22.25 &11.75 &1.5 &190.5 &21.84  & 8.86 & 1.3\cr
MRK\,915	  	&24.6  &22.57 &15.21 &1.1 &53.8  &22.76  &12.41 & 1.0\cr
UGC\,12138	  	&38.4  &24.38 &14.78 &2.2 &36.9  &22.21  &12.91 & 1.0\cr
AKN\,564	  	&32.7  &23.79 &14.82 &1.1 &39.7  &22.57  &13.16 & 0.9\cr
UGC\,12348	  	&30.7  &22.34 &15.78 &1.2 &38.4  &20.32  &13.25 & 1.2\cr
NGC\,7674	  	&27.7  &22.33 &14.62 &1.6 &49.2  &21.57  &12.25 & 1.2\cr
MRK\,590		&61.4  &24.23 &13.80 &2.5 &58.4  &21.91  &11.78 & 1.2\cr
MRK\,1058		&35.7  &23.94 &15.52 &2.1 &37.7  &22.19  &13.21 & 1.2\cr
NGC\,1386		&155.1 &23.67 &12.22 &2.1 &184.3 &22.49  &9.95  & 1.2\cr
MCG\,+8-11-11		&81.3  &24.10 &14.17 &1.0 &99.2  &22.21  &11.52 & 0.9\cr
MRK\,705		&34.4  &24.30 &14.89 &1.6 &44.0  &23.03  &12.99 & 1.6\cr
NGC\,3281		&178.2 &24.45 &12.82 &1.1 &184.3 &22.36  &10.44 & 1.2\cr
UGC\,6100		&48.2  &23.77 &14.66 &1.8 &60.5  &22.37  &12.70 & 1.9\cr
NGC\,4941		&190.5 &24.18 &12.30 &1.0 &192.0 &22.17  &10.16 & 1.1\cr
UGC\,10683B		&27.5  &23.31 &16.13 &1.3 &---   &---    &---   & ---\cr
ESO\,103-G35		&30.7  &22.48 &15.22 &1.4 &63.0  &22.29  &12.43 & 1.3\cr
NGC\,7212		&30.7  &22.60 &15.60 &1.5 &27.7  &20.14  &13.34 & 1.3\cr
\tablenotetext{}{Column 1 shows the galaxies names. Column 2 (6) presents the diameter of
the ellipse major axis used to measure the galaxy flux in B (I); Column 3 (7) presents
the surface brightness of the 3  $\sigma$ level above the background in B (I); Column 4 (8)
presents the integrated B (I) magnitude; and Column 5 (9) presents the seeing during
the observation.}
\enddata
\end{deluxetable}
\end{small}

\clearpage
\newpage

\begin{small}
\begin{deluxetable}{lrrrrrrrr}
\tablewidth{0pc}
\tablecaption{Magnitudes of the Additional Sample}
\tablehead{\colhead{Name}&\colhead{Aperture B}&
\colhead{3$\sigma$B}&\colhead{B}&\colhead{(seeing B)}&\colhead{Aperture I}& 
\colhead{3$\sigma$I}&\colhead{I}&\colhead{(seeing I)}\cr
&(arcsec)&(mag arcsec$^{-2}$)&(mag)&(arcsec)&
(arcsec)&(mag arcsec$^{-2}$)&(mag)&(arcsec)}
\startdata
ESO\,362-G8		&63.0  &24.34 &13.36 &1.4 &70.7  &22.67  &11.45 & 1.0\cr
NGC\,2273		&88.1  &23.54 &13.05 &1.9 &144.5 &22.53  &10.60 & 2.0\cr
ESO\,428-G14		&89.1  &24.32 &13.47 &1.2 &107.5 &21.84  &10.83 & 1.0\cr
MRK\,78			&24.8  &23.83 &15.56 &2.2 &33.0  &23.41  &13.51 & 1.4\cr
NGC\,2622		&30.3  &22.67 &15.17 &1.8 &37.2  &21.55  &12.86 & 1.7\cr
NGC\,2639		&77.1  &23.23 &12.92 &1.7 &114.2 &22.46  &10.55 & 2.2\cr
UGC\,5101		& ---  & ---  &  --- &--- &48.2  &23.03  &13.36 & 1.7\cr
NGC\,2992		& ---  & ---  &  --- &--- &115.2 &22.25  &10.96 & 1.1\cr
NGC\,3227		& ---  & ---  &  --- &--- &242.2 &22.12  &9.61  & 1.7\cr
MRK\,34			& ---  & ---  &  --- &--- &27.5  &21.81  &13.67 & 1.9\cr
NGC\,3362		& ---  & ---  &  --- &--- &59.2  &22.42  &12.32 & 1.8\cr
MRK\,423		& ---  & ---  &  --- &--- &34.4  &21.09  &13.18 & 2.1\cr
NGC\,5252		& ---  & ---  &  --- &--- &74.3  &22.58  &12.03 & 1.5\cr
MRK\,268		& ---  & ---  &  --- &--- &48.2  &23.11  &13.04 & 1.3\cr
MRK\,270		&46.8  &23.95 &14.38 &1.4 &74.3  &23.12  &12.00 & 1.2\cr
NGC\,5273		&114.2 &23.86 &12.70 &1.3 &151.4 &22.88  &10.52 & 1.7\cr
IC\,4329A		& ---  & ---  &  --- &--- &84.5  &22.64  &11.45 & 1.6\cr
NGC\,5929		&55.0  &22.74 &14.02 &1.3 &41.3  &20.54  &12.23 & 1.2\cr
NGC\,7450		& ---  & ---  &  --- &--- &45.6  &21.93  &12.99 & 1.4\cr
NGC\,7672		& ---  & ---  &  --- &--- &48.6  &22.52  &13.00 & 1.6\cr
\tablenotetext{}{Column 1 gives the galaxies names. Column 2 (6)
presents the diameter of the ellipse major axis used to measure the
galaxy flux in B (I); Column 3 (7) presents the surface brightness of
the 3  $\sigma$ level above the background in B (I); Column 4 (8)
presents the integrated B (I) magnitude; and Column 5 (9) presents the
seeing during the observation.}
\enddata
\end{deluxetable}
\end{small}

\clearpage
\newpage

\begin{small}
\begin{deluxetable}{lrrrr}
\tablewidth{0pc}
\tablecaption{PA's and Ellipticities of the 60$\mu$m Sample}
\tablehead{\colhead{Name}&\colhead{PA$_{MA}$-I}&\colhead{e-I}&\colhead{PA$_{MA}$-B}&\colhead{e-B}\cr
&(degrees)&&(degrees)&}
\startdata
MRK\,348		&-45 &0.061 &-60 &0.066 \cr
TOL\,0109$-$38		&61  &0.555 &61  &0.507 \cr
MRK\,1    		&70  &0.361 &67  &0.347 \cr
MRK\,359  		&8   &0.167 &-21 &0.119 \cr
MRK\,573  		&58  &0.137 &4   &0.137 \cr
IRAS\,01475$-$0740 	&-28 &0.186 &-35 &0.161 \cr
ESO\,153$-$G20 		&8   &0.298 &-50 &0.465 \cr
MRK\,1040		&75  &0.702 &--- &---   \cr
ESO\,355$-$G25 		&-41 &0.086 &-29 &0.076 \cr
UGC\,2024 		&-30 &0.373 &-29 &0.342 \cr
NGC\,1068 		&-67 &0.106 &-5  &0.196 \cr
MCG$-$02$-$08$-$039 	&12  &0.434 &17  &0.451 \cr
UGC\,2514   		&-40 &0.428 &-38 &0.309 \cr
IRAS\,03106$-$0254 	&-14 &0.681 &-17 &0.585 \cr
IRAS\,03125$+$0119 	&-32 &0.439 &-29 &0.440 \cr
MRK\,607 		&-42 &0.616 &-42 &0.630 \cr
ESO\,116$-$G18		&--- &---   &-87 &0.567 \cr
IRAS\, 04385$-$0828	&44  &0.517 &44  &0.493 \cr
IRAS\, 04502$-$0317	&15  &0.489 &13  &0.510 \cr
IRAS\, 04507$+$0358	&46  &0.192 &52  &0.310 \cr
ESO\,33$-$G02 		&-88 &0.086 &-74 &0.069 \cr
MCG$-$05$-$13$-$017 	&-41 &0.193 &-40 &0.199 \cr
MRK\,3	 		&28  &0.159 &26  &0.223 \cr
UGC\,3478     		&44  &0.623 &43  &0.518 \cr
MRK\,6	 		&-46 &0.411 &-48 &0.379 \cr
FAIRALL\,265		&2   &0.435 &2   &0.460 \cr
MRK\,79	  		&-47 &0.186 &69  &0.393 \cr
MRK\,10	   		&-53 &0.494 &-44 &0.513 \cr
UGC\,4155	   	&-40 &0.231 &-75 &0.312 \cr
MRK\,622	   	&-35 &0.098 &-46 &0.089 \cr
ESO\,18$-$G09		&-3  &0.138 &-22 &0.140 \cr
MCG$-$01$-$24$-$012	&42  &0.414 &38  &0.527 \cr
MRK\,1239 		&-25 &0.188 &-24 &0.193 \cr
NGC\,3393		&41  &0.131 &48  &0.128 \cr
NGC\,3516 		&56  &0.191 &56  &0.187 \cr
IRAS\,11215$-$2806 	&-34 &0.659 &-34 &0.649 \cr
MCG$-$05$-$27$-$013	&-82 &0.640 &-82 &0.663 \cr
MRK\,176     		&59  &0.609 &59  &0.612 \cr
NGC\,3783     	 	&-46 &0.074 &-44 &0.155 \cr
MRK\,766    		&67  &0.152 &70  &0.144 \cr
NGC\,4388   		&-89 &0.665 &-88 &0.722 \cr
NGC\,4507	    	&60  &0.164 &65  &0.122 \cr
NGC\,4593	  	&-72 &0.295 &66  &0.328 \cr
TOL\,1238$-$364    	&72  &0.104 &81  &0.078 \cr
NGC\,4704	    	&-61 &0.083 &-84 &0.037 \cr
MCG$-$02$-$33$-$034	&--- &---   &--- &---   \cr
ESO\,323$-$G32 		&-89 &0.097 &-70 &0.091 \cr
MCG$-$04$-$31$-$030	&58  &0.531 &58  &0.529 \cr
IRAS\,13059$-$2407 	&-59 &0.669 &-59 &0.688 \cr
MCG$-$03$-$34$-$064	&48  &0.318 &47  &0.231 \cr
ESO\,383$-$G18		&88  &0.546 &88  &0.552 \cr
MCG$-$6$-$30$-$15	&-76 &0.352 &-74 &0.359 \cr
NGC\,5347	 	&-60 &0.211 &-62 &0.223 \cr
IRAS\,14082$+$1347	&-85 &0.317 &-83 &0.305 \cr
NGC\,5506	   	&-89 &0.755 &-89 &0.758 \cr
NGC\,5548	   	&-80 &0.143 &-87 &0.198 \cr
IRAS\,14317$-$3237	&3   &0.245 &-1  &0.249 \cr
IRAS\,14434$+$2714	&88  &0.120 &77  &0.081 \cr
UGC\,9826	   	&-9  &0.327 &7   &0.300 \cr
UGC\,9944	   	&-7  &0.669 &-7  &0.676 \cr
IRAS\,15480$-$0344	&49  &0.113 &48  &0.066 \cr
IRAS\,16288$+$3929	&-77 &0.572 &-77 &0.568 \cr
IRAS\,16382$-$0613	&--- &---   &89  &0.253 \cr
UGC\,10889 	  	&43  &0.580 &45  &0.598 \cr
MCG$+$03$-$45$-$003	&-33 &0.428 &11  &0.302 \cr
FAIRALL\,49	  	&-59 &0.111 &-28 &0.186 \cr
FAIRALL\,51	  	&-22 &0.447 &2   &0.576 \cr
ESO\,143$-$G09		&35  &0.332 &34  &0.351 \tablebreak
FAIRALL\,341		&-23 &0.235 &-26 &0.228 \cr
UGC\,11630	  	&-81 &0.530 &-79 &0.436 \cr
PKS\,2048$-$57		&-60 &0.156 &-62 &0.167 \cr
NGC\,7213	  	&48  &0.033 &75  &0.026 \cr
MRK\,915	  	&-7  &0.507 &-18 &0.579 \cr
UGC\,12138	  	&-2  &0.077 &69  &0.024 \cr
AKN\,564	  	&-73 &0.198 &-72 &0.175 \cr
UGC\,12348	  	&-46 &0.715 &-46 &0.708 \cr
NGC\,7674	  	&-26 &0.266 &76  &0.228 \cr
MRK\,590		&-55 &0.087 &-73 &0.068 \cr
MRK\,1058		&-68 &0.418 &-68 &0.422 \cr
NGC\,1386		&25  &0.628 &25  &0.638 \cr
MCG\,+8-11-11		&54  &0.190 &1   &0.514 \cr
MRK\,705		&77  &0.226 &80  &0.225 \cr
NGC\,3281		&-39 &0.589 &-37 &0.599 \cr
UGC\,6100		&17  &0.336 &21  &0.360 \cr
NGC\,4941		&17  &0.518 &16  &0.529 \cr
UGC\,10683B		&--- &---   &--- &---  \cr
ESO\,103-G35		&39  &0.597 &40  &0.610 \cr
NGC\,7212		&43  &0.638 &42  &0.621 \cr
\tablenotetext{}{Column 1 gives the galaxy name; Columns 2 and 3 give
the position angle (PA$_{MA}$) and the ellipticity of the major axis
(e=1-b/a), and are also the values used to measure the I magnitudes;
Columns 4 and 5 give the position angle and ellipticity of the ellipse
used to measure the B magnitude, presented in Table 4.}
\enddata
\end{deluxetable}
\end{small}

\clearpage
\newpage

\begin{small}
\begin{deluxetable}{lrrrr}
\tablewidth{0pc}
\tablecaption{PA's and Ellipticities of the Additional Sample}
\tablehead{\colhead{Name}&\colhead{PA$_{MA}$-I}&\colhead{e-I}&\colhead{PA$_{MA}$-B}&\colhead{e-B}\cr
&(degrees)&&(degrees)&}
\startdata
ESO\,362-G8		&-15 &0.543 &-16 &0.537 \cr
NGC\,2273		&62  &0.362 &-87 &0.356 \cr
ESO\,428-G14		&-44 &0.343 &-39 &0.313 \cr
MRK\,78			&84  &0.435 &84  &0.430 \cr
NGC\,2622		&52  &0.332 &52  &0.329 \cr
NGC\,2639		&-43 &0.398 &-43 &0.455 \cr
UGC\,5101		&79  &0.417 &--- &---   \cr
NGC\,2992		&13  &0.556 &--- &---   \cr
NGC\,3227		&-31 &0.459 &--- &---   \cr
MRK\,34			&65  &0.130 &--- &---   \cr
NGC\,3362		&0   &0.062 &--- &---   \cr
MRK\,423		&--- &---   &--- &---   \cr
NGC\,5252		&15  &0.480 &--- &---   \cr
MRK\,268		&-67 &0.415 &--- &---   \cr
MRK\,270		&-60 &0.137 &-61 &0.156 \cr
NGC\,5273		&11  &0.156 &10  &0.170 \cr
IC\,4329A		&44  &0.464 &--- &---   \cr
NGC\,5929		&49  &0.423 &48  &0.437 \cr
NGC\,7450		&25  &0.321 &--- &---   \cr
NGC\,7672		&28  &0.281 &--- &---   \cr
\tablenotetext{}{Column 1 gives the galaxy name; Columns 2 and 3 give
the position angle (PA$_{MA}$) and the ellipticity of the major axis
(e=1-b/a), and are also the values used to measure the I magnitudes;
Columns 4 and 5 give the position angle and ellipticity of the ellipse
used to measure the B magnitude, presented in Table 5.}
\enddata
\end{deluxetable}
\end{small}